\documentclass[conference]{IEEEtran}
\IEEEoverridecommandlockouts
\usepackage{cite}
\usepackage{amsmath,amssymb,amsfonts}
\usepackage{algorithmic}
\usepackage{graphicx}
\usepackage{textcomp}
\usepackage{xcolor}
\def\BibTeX{{\rm B\kern-.05em{\sc i\kern-.025em b}\kern-.08em
    T\kern-.1667em\lower.7ex\hbox{E}\kern-.125emX}}

\usepackage{geometry}
\begin{document}

\title{Decentralization Issues in Cell-free Massive MIMO Networks with Zero-Forcing Precoding\\
\thanks{Work supported by MINECO under project TERESA-TEC2017-90093-C3-3-R (AEI/FEDER,UE), Spain.}
}

\author{\IEEEauthorblockN{Felip Riera-Palou and Guillem Femenias}
\IEEEauthorblockA{\textit{Mobile Communications Group - University of the Balearic Islands}, Mallorca (Illes Balears), Spain \\
\{felip.riera,guillem.femenias\}@uib.es}
}

\maketitle

\begin{abstract}
 Cell-free massive MIMO (CF-M-MIMO) systems represent an evolution of the classical cellular architecture that has dominated the mobile landscape for decades. In CF-M-MIMO, a central processing unit (CPU) controls a multitude of access points (APs) that are irregularly scattered throughout the coverage area effectively becoming a fully distributed implementation of the M-MIMO technology. As such, it inherits many of the key properties that have made M-MIMO one of the physical layer pillars of 5G systems while opening the door to new features not available in M-MIMO. Among the latest is the possibility of performing the precoding at the CPU (centralized) or at the APs (distributed) with the former known to offer much better performance at the cost of having to collect all the relevant channel state information (CSI) at the CPU. Realistic deployments of cell-free systems are likely to require more than one CPU when the area to be covered is large, thus a critical issue that needs to be solved is how these multiple CPUs should be interconnected. This paper analyzes and proposes designs for different degrees of interconnectivity among the CPUs for the specific case of centralized zero-forcing precoding. Results show that a modest form of CPU interconnection can boost very significantly the max-min rate performance so prevalent in CF-M-MIMO architectures.
\end{abstract}


\section{Introduction}
The evergrowing mobile user expectations in terms of throughput, coverage and ubiquity are pushing the wireless research community to investigate novel network architectures well beyond the present incarnation of 5G currently being deployed worldwide. Recently, the concept of cell-free Massive-MIMO (CF-M-MIMO) has received considerable interest mainly due to its inherent ability to provide a uniform quality-of-service (QoS) throughout the deployment area. Initially proposed in \cite{ngo15,ngo17}, CF-M-MIMO assumes the existence of a single central processing unit (CPU) to which a plethora of access points (APs), irregularly distributed over the area to be covered, are connected via fronthaul links. CF-M-MIMO has been shown to allow a distributed implementation of a conventional M-MIMO system by relying on a conjugate beamforming (CB) precoder, locally computed at each AP, while centrally optimizing the power allocation mechanism. This strategy allows all users in the network to attain the same throughput (max-min optimization). Interestingly, this centralized power optimization only relies on large-scale channel state information (CSI), thus greatly alleviating throughput and latency constraints on the fronthaul links. Almost at the same time, the performance of the CF-M-MIMO when using zero-forcing (ZF) precoding was studied in \cite{nayebi17}, showing that it greatly outperforms CB precoding in terms of max-min rate although at the cost of having to centralize the precoder design at the CPU, a procedure that requires of short-term CSI and therefore poses stronger requirements on the fronthaul \cite{femenias19}.

When transiting from theoretical proposals to practical deployments, an important issue that should be confronted is whether just a single CPU can control all the APs in the coverage area. In most practical situations, when the area to be serviced expands, multiple CPUs will need to be deployed, possibly connected by backhaul links. Very recently, \cite{interdonato19} has touched upon the scalability aspects of cell-free systems by considering the use of multiple CPUs, each controlling a set of APs, that jointly serve the users in the area. In their approach, groups of users and APs are formed so that a given user will typically be served by the group of APs/CPUs located nearby, an idea reminiscent of the user-centric approach proposed in \cite{buzzi17} or the clustering technique introduced in \cite{riera-palou18b}. Authors in \cite{interdonato19} show that independently operated CPUs, when using CB precoding, do not loose much in terms of spectral efficiency with respect to a fully centralized system. However, in their work, the power optimization step was not designed to ensure the uniform rate for all users in the network. In this work we generalize some aspects of \cite{interdonato19} by considering different degrees of interconnectivity among the CPUs (from unconnected to strongly connected) while focusing on the performance achieved when using ZF precoding. Moreover, a power allocation strategy is proposed that \newgeometry{left=19.1mm, top=19.1mm}\noindent still guarantees the uniform achievable rate property for all users in the network while having modest requirements in terms of CPU interconnection capabilities (i.e., backhaul requirements).

\section{System model}
Let us start by considering a conventional cell-free architecture along the lines described in the seminal papers \cite{ngo17,nayebi17} and that will be used here as a reference baseline. In particular, this system is formed by $M$ single-antenna APs, each with available transmit power $P_t^{\text{AP}}$, connected by means of fronthaul links to a CPU and is in charge of serving $K$ single-antenna mobile stations (MSs). In this work it is assumed that the fronthaul links have unlimited capacity but it is worth stating at this point that expanding the current submission to take into account finite-capacity fronthaul links constitutes a promising avenue for further research. For brevity of exhibition, this paper focuses on the downlink, notwithstanding the fact that most of the discussion also applies to the uplink segment. Unlike \cite{interdonato19}, where a distributed precoding strategy was examined (i.e., conjugate beamforming implemented at the APs), this work considers the use of ZF precoding implemented at the CPU.

As it is typically done in M-MIMO, downlink and uplink transmissions are organized in a time division duplex (TDD) operation whereby each coherence interval is split into three phases, namely, the uplink training phase, the downlink payload data transmission phase and the uplink payload data transmission phase. In the uplink training phase, all MSs transmit uplink training pilots allowing the AP to estimate the propagation channels to every MS in the network\footnote{Note that channel reciprocity can be exploited in TDD systems and therefore only uplink pilots need to be transmitted.}. Subsequently, these channel estimates are used to compute the precoding filters governing the downlink payload data transmission and to detect the signals transmitted from the MSs in the uplink payload data transmission phase. Critically, the combined duration of the training, downlink and uplink phases, denoted as $\tau_p$, $\tau_\text{dl}$ and $\tau_\text{ul}$, respectively, should not exceed the coherence time of the channel, denoted as $\tau_c$, that is, $\tau_p+\tau_\text{dl}+\tau_\text{ul}\leq \tau_c$, with all these times specified sampling periods.

The propagation channel linking AP $m$ to MS $k$ is denoted by $g_{mk}$ and modelled as
\begin{equation}
   g_{mk}=\sqrt{\beta_{mk}} h_{mk},
\end{equation}
where $\beta_{mk}$ represents the large-scale propagation losses (i.e., path loss and shadowing) and $h_{mk}$ correspond to the small-scale fading coefficient. The large-scale gain is further decomposed as $\beta_{mk}=\zeta_{mk}\chi_{mk}$ with $\zeta_{mk}$ representing the distance-dependent path loss and $\chi_{mk}$ corresponding to the shadowing component. For comparative purposes, in this work use is made of exactly the same large-scale losses model introduced in \cite{ngo17}. In particular, $\zeta_{mk} \ \forall\,mk$ adheres to the three-slope path loss model described in \cite[(52)-(53)]{ngo17} while the shadowing component $\chi_{mk}$ is modelled as a correlated log-normal random variable with variance $\sigma_\chi^2$ whose spatial correlation model is described in \cite[(54)-(55)]{ngo17}. Finally, the small-scale fading terms $h_{mk}$ consist of independent and identically distributed (i.i.d.) complex Gaussian random variables distributed as $\mathcal{CN}(0,1)$. The channel coefficients $g_{mk}$ are assumed to be static throughout the coherence interval and then change independently (i.e., block fading). As in the seminal papers \cite{ngo15,ngo17} introducing the idea of cell-free operation, it is assumed that the CPU has perfect knowledge of the large-scale fading gains (i.e., $\beta_{mk}$ $\forall\,mk$) and, if required, can make them available to the different APs.

Communication in any coherence interval of a TDD-based M-MIMO system invariably starts with the MSs sending the pilot sequences to allow the channel to be estimated at the AP. Channel estimation is known to play a central role in the performance of M-MIMO schemes \cite{lu14} and also in the specific context of cell-free architectures \cite{buzzi17}. During the uplink training phase, all $K$ MSs simultaneously transmit pilot sequences of $\tau_p$ samples to the APs and thus, the $\tau_p\times 1$ received uplink signal at the $m$th AP is given by
\begin{equation}
   {\boldsymbol{y}_p}_m=\sqrt{\tau_p P_t^\text{MS}}\sum_{k=1}^K g_{mk}\boldsymbol{\varphi}_k+{\boldsymbol{w}_p}_m,
\end{equation}
where $P_t^\text{MS}$ is the transmit power available at the MSs for pilot symbol transmission, $\boldsymbol{\varphi}_k$, with $\|\boldsymbol{\varphi}_k\|^2=1$, denotes the $\tau_p\times 1$ training sequence assigned to MS $k$ and ${\boldsymbol{w}_p}_m$ is a $\tau_p\times 1$ vector of i.i.d. additive noise samples with each entry distributed as $\mathcal{CN}(0,\sigma_\upsilon^2)$. Ideally, training sequences should be chosen to be mutually orthogonal, however, since in most practical scenarios it holds that $K>\tau_p$, a given training sequence is assigned to more than one MS, thus resulting in the so-called pilot contamination, a widely studied phenomenon in the context of centralized M-MIMO systems \cite{Elijah16}. In this work it is assumed that training sequences are assigned to MSs using the fingerprinting technique introduced in \cite{femenias19}. This strategy ensures that pilot sequences are reused only among users which are located far apart from each other, hence reducing the pilot contamination effects (see \cite[Section V]{femenias19} for details).

Given ${\boldsymbol{y}_p}_m$, the MMSE estimate of $g_{mk}$ can be calculated as \cite{ngo17}
\begin{equation}
   \hat{g}_{mk}=\frac{\sqrt{\tau_p P_t^\text{MS}}\beta_{mk}}{\xi_{mk}}\boldsymbol \varphi_k^H  {\boldsymbol{y}_p}_m,
\end{equation}
where
\begin{equation}
   \xi_{mk}=\tau_pP_t^\text{MS}\sum_{k'=1}^K \beta_{mk'}\,\left|\boldsymbol{\varphi}_{k'}^H\boldsymbol{\varphi}_k\right|^2+\sigma_\upsilon^2.
\end{equation}
For notational convenience, the $K\times 1$ vector collecting the channel responses from AP $m$ to all $K$ MSs in the network is defined as $\boldsymbol g_k=\left[g_{1k} \ldots g_{Mk} \right]^T$ and its corresponding estimate as $\hat{\boldsymbol g}_k$. Similarly, the $M\times K$ matrix $\mathbf G=\left[\boldsymbol g_1 \ldots \boldsymbol g_K \right]$ collects the channel responses between the $M$ APs and the $K$ MSs and $\hat{\mathbf G}$ correspond to its MMSE estimate. Finally, the channel estimation error will be denoted by $\tilde{g}_{mk}={g}_{mk}-\hat{g}_{mk}$ and in vector/matrix forms as $\tilde{\boldsymbol g}_k$ and $\tilde{\mathbf G}$.

\section{Max-Min Zero-Forcing (MM-ZF) precoding}
At each signaling interval, the CPU processes a $K\times 1$ vector $\boldsymbol s=\left[s_1 \ldots s_K \right]^T$ of information symbols with $E\{|s_k|^2\}=1$ using linear precoding as
\begin{equation}
\boldsymbol x = \left[x_1 \ldots x_M \right]^T = \mathbf W \mathbf P \boldsymbol s,
\end{equation}
where the $K\times M$ matrix $\mathbf W$ corresponds to the ZF precoding operation that is given by
\begin{equation}
\mathbf W=\hat{\mathbf G}^H \left(\hat{\mathbf G} \hat{\mathbf G}^H \right)^{-1},
\end{equation}
and $\mathbf P=\text{diag}([\eta_1^{1/2} \ldots \eta_K^{1/2}])$ is a power allocation diagonal matrix\footnote{$\mathcal D\left(\mathbf A\right)$ denotes the vector formed by the main diagonal of matrix $\mathbf A$ whereas $\text{diag}(\boldsymbol x)$ denotes a diagonal matrix with vector $\boldsymbol x$ at its main diagonal.} with $\eta_k$ denoting the power coefficient applied to user $k$.
The estimated symbol by an arbitrary MS $k$, $\hat s_k$, is then given by
\begin{equation}
\hat s_k=\mathbf g_k^T \boldsymbol x+\upsilon_k,
\end{equation}
with $\upsilon_k\sim \mathcal C\mathcal N (0,\sigma_\upsilon^2)$ denoting the receiver additive white Gaussian noise (AWGN) sample. A tight lower bound to the signal-to-noise-plus-interference ratio for user $k$ can then be expressed as \cite{nayebi17}
\begin{equation}
\texttt{SINR}_k=\frac{P_t^\text{AP}\eta_k}{P_t^\text{AP}\sum_{k'=1}^{K}\gamma_{kk'}\eta_{k'}^{1/2}+\sigma_\upsilon^2},
\label{SINR}
\end{equation}
where $\gamma_{kk'}$ is the $k'$th entry in vector
\begin{equation}
\small
\boldsymbol \gamma_k=\mathcal D \left(E\left\{\left(\hat{\mathbf G}^T\hat{\mathbf G}^*\right)^{-1} \hat{\mathbf G}^T E\{\tilde {\mathbf g}_k^*\tilde {\mathbf g}_k^T \} \hat{\mathbf G}^*  \left(\hat{\mathbf G}^T \hat{\mathbf G}^* \right)^{-1} \right\}\right)
\label{gamma1}
\end{equation}
with $E\{\tilde {\mathbf g}_k^*\tilde {\mathbf g}_k^T \}=\text{diag}\left([\beta_{1k}-\alpha_{1k}, \ldots, \beta_{Mk}-\alpha_{Mk}] \right)$ and $\alpha_{mk}=\frac{P_t^\text{AP}\tau_p\beta_{mk}^2}{\sigma_\upsilon^2+P_t^\text{AP}\tau_p\beta_{mk}}$. Note that $\boldsymbol \gamma_k$ represents the vector of interfering terms caused by the use of the channel estimates, rather than the true values, when designing the ZF matrix $\mathbf W$. As in \cite{nayebi17}, it is worth mentioning that the outer expectation in \eqref{gamma1} cannot be computed in closed-form and must be estimated via Monte-Carlo simulation. Using \eqref{SINR}, an achievable rate for user $k$ can be defined as $R_k=\log_2(1+\texttt{SINR}_k)$.

In line with the original cell-free philosophy, the power coefficients $\eta_k$ are chosen to maximize the minimum $\texttt{SINR}_k$ for $k \in \{1,\ldots,K\}$, subject to a maximum power constraint at each AP. Mathematically,
\begin{equation}
\begin{split}
&\substack{\max\min\\\eta_1,\ldots,\eta_K} \texttt{ SINR}_k \ \\
&\text{s.t. } |x_m|^2\leq P_t^\text{AP} \ \ \forall m \in \{1,\ldots,M\}.
\label{optim}
\end{split}
\end{equation}
The maximization in  \eqref{optim} has been shown to be a quasi-linear convex problem that can be effectively solved using the bisection method \cite{nayebi17}.
\begin{figure}[t!]
    \centering
    \includegraphics[width=.75\linewidth,angle=-90]{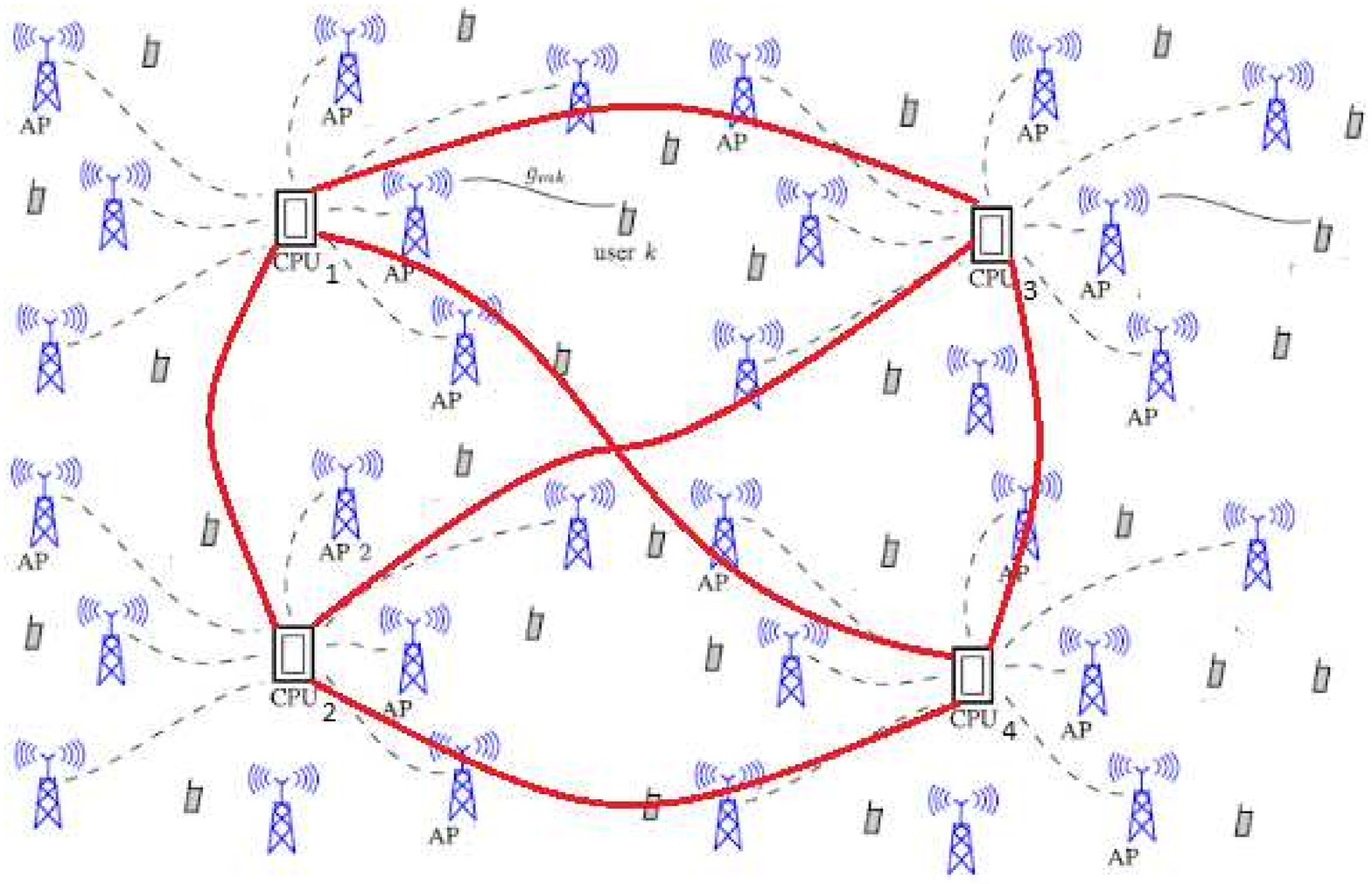}
    \caption{(Modified from \cite{ngo17}) Partitioned CF-M-MIMO deployment with $D=4$ CPUs. Thick Lines (in red) represent the backhaul lines connecting the different CPUs. Note that the connection from each CPU to every other CPU need not necessarily be enforced as long as each CPU can reach any other CPU through multiple hops.}
    \label{fig.system}
\end{figure}

\section{Decentralization strategies}
Following the argument introduced in \cite{interdonato19}, the existence of a unique CPU to govern the whole coverage area is bound to become unpractical as the scale of the area to be covered expands, thus requiring, at some point, multiple CPUs to be deployed. Towards this end, we envisage an scenario such as the one shown in Fig.~\ref{fig.system} where several CPUs are scattered throughout the coverage area, each of them controlling a multitude of APs. As stated in \cite{interdonato19}, it is assumed that the CPUs are linked so that at least they are able to transmit synchronously. The main issue addressed in this paper is to propose designs for different degrees of interconnectivity (i.e. different backhaul requirements) among the CPUs while assessing the performance each of them offers. Prior to that, and taking the single CPU scenarios as a baseline, any decentralization strategy begins by considering how the association from APs to CPUs and from users to APs is conducted.

\subsection{Clusterization and user association}
Towards this end, we assume that the $M$ APs are clustered using the procedure described in \cite{riera-palou18b}. This strategy relies on applying the k-means algorithm using the Euclidean distance as clustering metric and assuming the AP positions are known. Subsequently, each user is allocated to the cluster whose average large-scale losses from the APs conforming the cluster to the considered user are minimum, thus ensuring that each user is served by the cluster (i.e., CPU) with minimum average large-scale losses. Note that whereas the AP clustering has to be considered a one-off procedure conducted during network planning, the MS clustering is an on-going process that will typically be conducted on the large-scale gain time-scale, thus it is able to capture the mobility patterns different users may have. As a result of this clustering strategy, and allowing the possibility of different CPUs controlling different parts of the network, it is assumed from this point onwards the existence of $D$ distinct CPUs, denoting by $M_d$ and $K_d$ the number of APs and users, respectively, served by CPU $d$.
In this paper, our focus is restricted to disjoint clusters of APs and to user-association strategies where each user is exclusively assigned to a single CPU. The relaxations of these two conditions constitute interesting topics for future research. Note that each CPU has access only to the data of the users it is serving.

Building on the clusterization step, let us denote by $\hat{\mathbf G}^{(d)}$ the $M_d\times K_d$ matrix collecting the estimated channel coefficients from the APs controlled by CPU $d$ to the $K_d$ users assigned to the cluster controled by CPU $d$. Similarly, define ${\overline{\mathbf G}}^{(d)}$ as the $M_d\times K$ estimated channel matrix between the APs controlled by CPU $d$ and all the users in the coverage area\footnote{Note pilot contamination affecting the uplink training can arise from users controlled by any CPU including the $d$th one.}. The transmit signal from CPU $d$ is now given by
\begin{equation}
\boldsymbol x^{(d)} = \mathbf W^{(d)} \mathbf P^{(d)} \boldsymbol s^{(d)} \ \ \ \forall d \in \{1,\ldots,D\},
\end{equation}
where $\mathbf W^{(d)}, \mathbf P^{(d)} \text{and } \boldsymbol s^{(d)}$ correspond to the ZF precoder, power allocation matrix and vector of transmitted symbols at the $d$th CPU.
Now, the estimated symbol for an arbitrary user $k$ controlled by CPU $d$ follows as
\begin{equation}
\hat s_k^{d}=\underbrace{\left(\mathbf g_k^{(d)}\right)^T \boldsymbol x^{(d)}}_{T_0}+\underbrace{\sum_{d'\neq d}^{D}\mathbf g_k^{(d')} \boldsymbol x^{(d')}}_{T_1}+\upsilon_k,
\label{recep}
\end{equation}
with $\mathbf g_k^{(d)}$ denoting the channel coefficients from the APs controlled by CPU $d$ to user $k$. The first term ($T_0$) corresponds to the signal received from the desired CPU while the second term ($T_1$) collects the inter-CPU interference. Let us now consider the form of $\mathbf W^{(d)}$ and $\mathbf P^{(d)}$ when the CF-F-MIMO system is decentralized and how $\hat{\mathbf G}^{(d)}$ can be estimated in each case. In particular, we consider three decentralization strategies supported by different degrees of interconnection, namely,
\begin{itemize}
\item \textbf{Strong connectivity} (SC). Under this assumption the CPUs are able to exchange the locally collected short-term CSI (e.g., $\overline{\mathbf G}^{d}$ for $d=1,\ldots,D$) and thus enabling the calculation of an overall ZF precoding matrix acting upon all users' data. Note that this solution would necessarily require of large-bandwidth low-latency backhauls interconnecting the different CPUs given the fast variations of the CSI and the large number of coefficients that need to be exchanged. As it will be shortly shown, this solution scheme is mathematically equivalent to a fully centralized solution.
\item \textbf{Weak connectivity} (WC). In this case, CPUs are only able to exchange large-scale information, thus $\beta_{mk}$ for all $m,k$ can be assumed to be known at the $D$ CPUs as well as any other large-scale information estimated at each cluster. Critically, each CPU designs a local ZF precoding matrix $\mathbf W^{(d)}$ unaware of the rest of CPUs. Albeit this more restrictive form of inter-CPU information sharing, this architecture still enables a joint calculation of the power coefficients and a global management of the pilot sequence allocation since both these processes just rely on large-scale channel statistics. Obviously, backhaul requirements are far more modest given the slow variations of the large-scale information.
\item \textbf{No connectivity} (NC). Finally, an scenario is considered where the CPUs operate with total independence except for the synchronization and neither short- nor long-term CSI information is exchanged among the CPUs. Critically, power allocation and uplink pilot sequence allocation are conducted independently on a per-CPU basis.
\end{itemize}
The next subsection describes how the MM-ZF precoder can be implemented in a decentralized fashion under the three different connectivity conditions. For clarity of presentation, Table~\ref{table1} summarizes what information is available at each CPU under each decentralization setup (some of the variables in the table are described over the next paragraphs).

\begin{table*}[t]
\caption{Information required at each CPU for different connectivity models}
\label{table1}\centering
  \begin{tabular}{|c|c|c|c|}
  \hline
    \textbf{Connectivy type} & \textbf{Known variables at arbitrary CPU $d$} & \textbf{Additional info} & \textbf{Synchronization signal}\\
    \hline
    SC & $\beta_{mk}, \boldsymbol\gamma_k$ $\forall m,k$, $\mathbf G$ & $K_1, \ldots, K_D$, Complete pilot assignment & Yes\\
    \hline
    WC & $\beta_{mk}$ $\forall m,k$, $\boldsymbol \gamma_k^{(d)}, \overline{\boldsymbol \gamma}_k^{(d')}$ $\forall d,d',k$, $\hat{\mathbf G}^{(d)}$ $\forall d$ & $K_1, \ldots, K_D$, Complete pilot assignment & Yes \\
    \hline
    NC & $\beta_{mk}, \boldsymbol \gamma_k$ $\forall m$, $\forall k \in \mathcal C^{(d)}$, $\hat{\mathbf G}^{(d)}$ & $ K_d$ & Yes \\
    \hline
  \end{tabular}
  \end{table*}

\subsection{Decentralized MM-ZF precoding}
\subsubsection{Strong connectivity}
The MM-ZF precoding under SC can be solved relying on the following lemma.\\
\noindent\textbf{\emph{Lemma}}: \emph{In the specific case of disjoint clusters, the SC strategy with clustering results in the same solution as the one obtained on the basis of \eqref{SINR} and \eqref{optim} without clustering.\\}
\textbf{\emph{Proof}}: Each CPU can estimate the channel from the $K_d$ users it controls but also from the $K-K_d$ users from the other clusters since user pilot allocation can also be assumed to be known. Owing to the SC strategy, this CSI can be communicated to the other CPUs, and each of them can then compute the overall MM-ZF precoder (i.e., matrices $\mathbf W$ and $\mathbf P$). Local precoding proceeds by constructing $\mathbf W^{(d)}$  and $\mathbf P^{(d)}$ by selecting $K_d$ user-specific columns of the overall power-controlled precoder $\mathbf W\mathbf P$ and apply it to the available user data at each CPU to generate $\boldsymbol x^{(d)}$.  Effectively this becomes equivalent to a totally centralized scheme although implemented in a distributed manner similar to that used in network MIMO setups \cite{riera-palou18}.\\

\subsubsection{Weak connectivity}
The WC strategy allows the implementation of a CPU-coordinated power and pilot allocation but limits the design of the ZF part of the precoder at each CPU to be based solely on the local information. Mathematically, the local ZF matrix follows $\mathbf W^{(d)}=({\hat{\mathbf G}^{(d)}})^H\left(\hat{\mathbf G}^{(d)}(\hat{\mathbf G}^{(d)})^H\right)^{-1}$.\\
\textbf{\emph{Theorem}}: \emph{Denoting by $\mathcal C^{(d)}$ and $\mathcal M(d)$ the set of users and APs, respectively, controlled by CPU $d$, the SINR for user $k$ controlled by CPU $d$ is given by}
\begin{equation}
\resizebox{1\hsize}{!}{$\texttt{SINR}_k^{(d)}=\frac{P_t^\text{AP}\eta_k^{(d)}}{\displaystyle P_t^\text{AP}\sum_{k' \in \mathcal C^{(d)}}\gamma_{kk'}^{(d)}\eta^{(d)}_{k'}+P_t^\text{AP}\sum_{d'\neq d}^{D}\sum_{l' \in \mathcal C^{(d')}}\bar{\gamma}_{kl'}^{(d')}\eta_{l'}^{(d')}+\sigma_\upsilon^2},$}
\label{SINR2}
\end{equation}
where, similar to \eqref{SINR}$, \gamma_{kk'}^{(d)}$ is the $k'$th entry of vector
\begin{equation*}
\boldsymbol \gamma_k^{(d)}=\mathcal D\left(E\left\{\mathbf W^{(d)^H} E\{\tilde {\mathbf g}_k^{(d)^*}\tilde {\mathbf g}_k^{(d)^T} \} \mathbf W^{(d)} \right\}\right)
\end{equation*}
with $ E\{\tilde {\mathbf g}_k^{(d)^*}\tilde {\mathbf g}_k^{(d)^T} \}=\text{diag}\left([\beta_{m^{(1)}k}-\alpha_{m^{(1)}k}, \ldots,\right.$ $\left.\beta_{m^{(Md)}k}-\alpha_{m^{(Md)}k}]\right)$ with $m^{(i)}$ denoting the indices in $\mathcal M(d)$, and $\bar{\gamma}_{kl'}^{(d')}$ is the  $l'$th entry in
\begin{equation*}
\bar{\boldsymbol \gamma}_k^{(d')}=\mathcal D\left(E\left\{\mathbf W^{(d')^H}  {\mathbf g}_k^{(d')^*} {\mathbf g}_k^{(d')^T} \mathbf W^{(d')} \right\}\right).
\end{equation*}

\noindent\textbf{\emph{Proof (sketch)}}: The numerator and the first term in the denominator in \eqref{SINR2} follow exactly from the proof of \eqref{SINR} (see \cite[Appendix C]{nayebi17} for details) simply substituting the overall estimated matrix $\hat{\mathbf G}$ by the desired cluster-specific matrix $\hat{\mathbf G}^{(d)}$. The second denominator term corresponds to the interference power generated by the CPUs other than the desired one, reflected by the second term in \eqref{recep} (e.g., $T_1$). Assuming the transmission of independent symbols with unit power it holds
\begin{equation*}
\begin{split}
&E\left\{|T_1|^2\right\}= \sum_{d'\neq d}^{D}E\left\{|\boldsymbol g_k^{(d')}\mathbf W^{(d')} \mathbf P^{(d')} \boldsymbol s^{(d')}|^2\right\}\\
&=\sum_{d'\neq d}^{D} P_t^\text{AP} \text{Tr}\left[\mathbf P^{(d')^2} E\left\{\mathbf W^{(d')^H} \boldsymbol g_k^{(d')^*}\boldsymbol g_k^{(d')^T}\mathbf W^{(d')} \right\} \right]\\
&=P_t^\text{AP}\sum_{d'\neq d}^{D} \sum_{l' \in \mathcal C^{(d')}} \bar{\gamma}_{kl'}^{(d')}\eta_{l'}^{(d')}.
\end{split}
\end{equation*}
Note that $\boldsymbol \gamma_k^{(d)}$,$\bar{\boldsymbol \gamma}_k^{(d')}$ constitute large-scale information and, as indicated in Table~\ref{table1}, under the WC connectivity model it can be exchanged among all CPUs.

Having an expression for the SINR of an arbitrary user under WC, it is now possible to pose the max-min problem as
\begin{equation*}
\begin{split}
&\substack{\max\min\\ \boldsymbol\eta^{(1)},\ldots,\boldsymbol\eta^{(D)}} \texttt{ SINR}^{(d)}_k \\
&\text{s.t. } |x_m|^2\leq P_t^\text{AP} \ \ \forall m \in \{1,\ldots, M\},
\end{split}
\label{optim2}
\end{equation*}
where $\boldsymbol\eta^{(d)}=\left[\eta_1^{(d)}, \ldots, \eta_{K_d}^{(d)} \right]$. This problem bears great structural resemblance to the original one in \eqref{optim}. In fact, it can be shown to also be quasi-linear and thus, amenable to be solved using the bisection method just taking care of relying on the adequate large-scale parameters (i.e., $\beta_{mk}-\alpha_{mk}$ or $\beta_{mk}$) depending on whether the interfering term comes from the desired CPU or from an interfering one. Importantly, the fingerprinting pilot allocation from \cite{femenias19}, which only relies on large-scale parameters, can also be implemented under the WC setup.

\subsubsection{No connectivity}
Finally, the NC setup assumes that CPUs are totally unconnected (yet synchronized). Note now that the SINR experienced by an arbitrary user $k$ associated to CPU $d$ is the same as in the WC case (given by \eqref{SINR2}), however, as now each CPU has only knowledge of the long-term statistics of the users  it controls, power optimization is conducted following \eqref{optim} locally and independently (i.e., every CPU ignores the presence of other CPUs). As it will be seen, this uncoordinated power allocation greatly harms overall max-min performance.

\section{Numerical results}
We consider an scenario characterized by an $L\times L$ squared area with $L=1000$ m, where $M=100$ APs, each with $P_t^\text{AP}=200$ mW available transmit power, have been randomly distributed according to a Uniform distribution. Users are also deployed randomly (uniform) throughout the coverage area and it is assumed that pilot transmission is conducted using power $P_t^\text{MS}=100$ mW. The wrap-around technique has been applied to the simulation setup to minimize boundary effects. A noise power spectral density of $-174$ dBm/Hz and receiver noise figure of $9$ dB are assumed. The network operates over a bandwidth of $W=20$ MHz and the channel model parameters are selected as in \cite{nayebi17}. Channel coherence is assumed to span $\tau_c=200$ time/frequency samples, out of which $\tau_p=15$ are consumed for uplink training, with the remaining time equally split between the downlink and uplink. As it has been mentioned before, pilot allocation is conducted either globally (SC and WC cases) or locally (on a CPU-basis for the NC case) using the fingerprinting technique introduced in \cite{femenias19}. The pilot matrix $\boldsymbol\Phi$ has been chosen to be a $\tau_p \times \tau_p$ discrete Fourier transform (DFT) matrix. Based on this general environment, the downlink max-min performance of all users in the network is analyzed under the assumption of $D=1,2,3$ and $4$ CPUs and the various connectivity strategies. The results shown here have been obtained by averaging over 200 random throws of APs and MSs, where for each throw, 1000 independent fast fading realizations have been conducted.
\begin{figure}[t!]
    \centering
    \includegraphics[width=\linewidth]{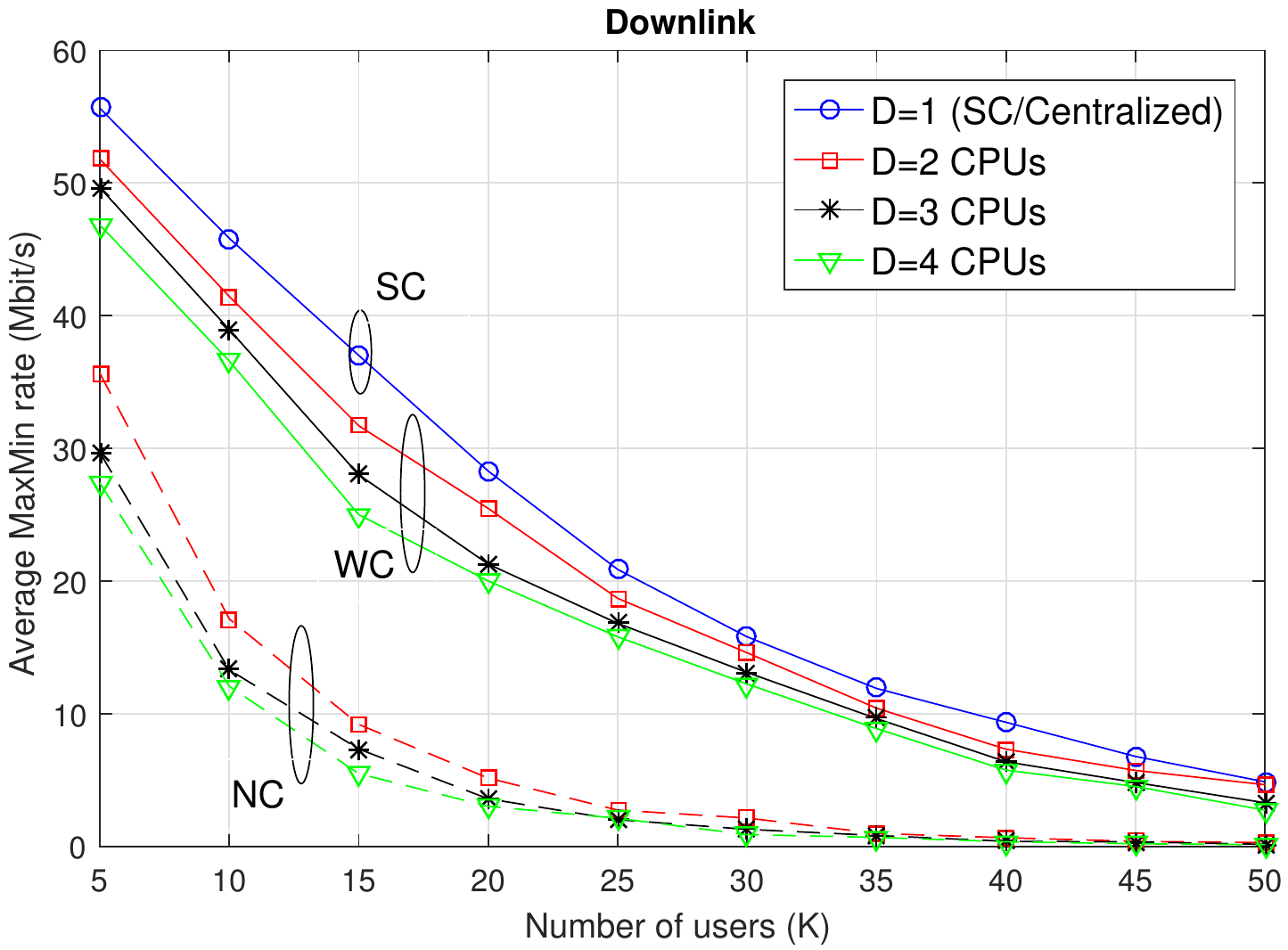}
    \caption{Average max-min performance for SC, WC and NC.}
    \label{max-min}
\end{figure}

\begin{figure}[t!]
    \centering
    \includegraphics[width=\linewidth]{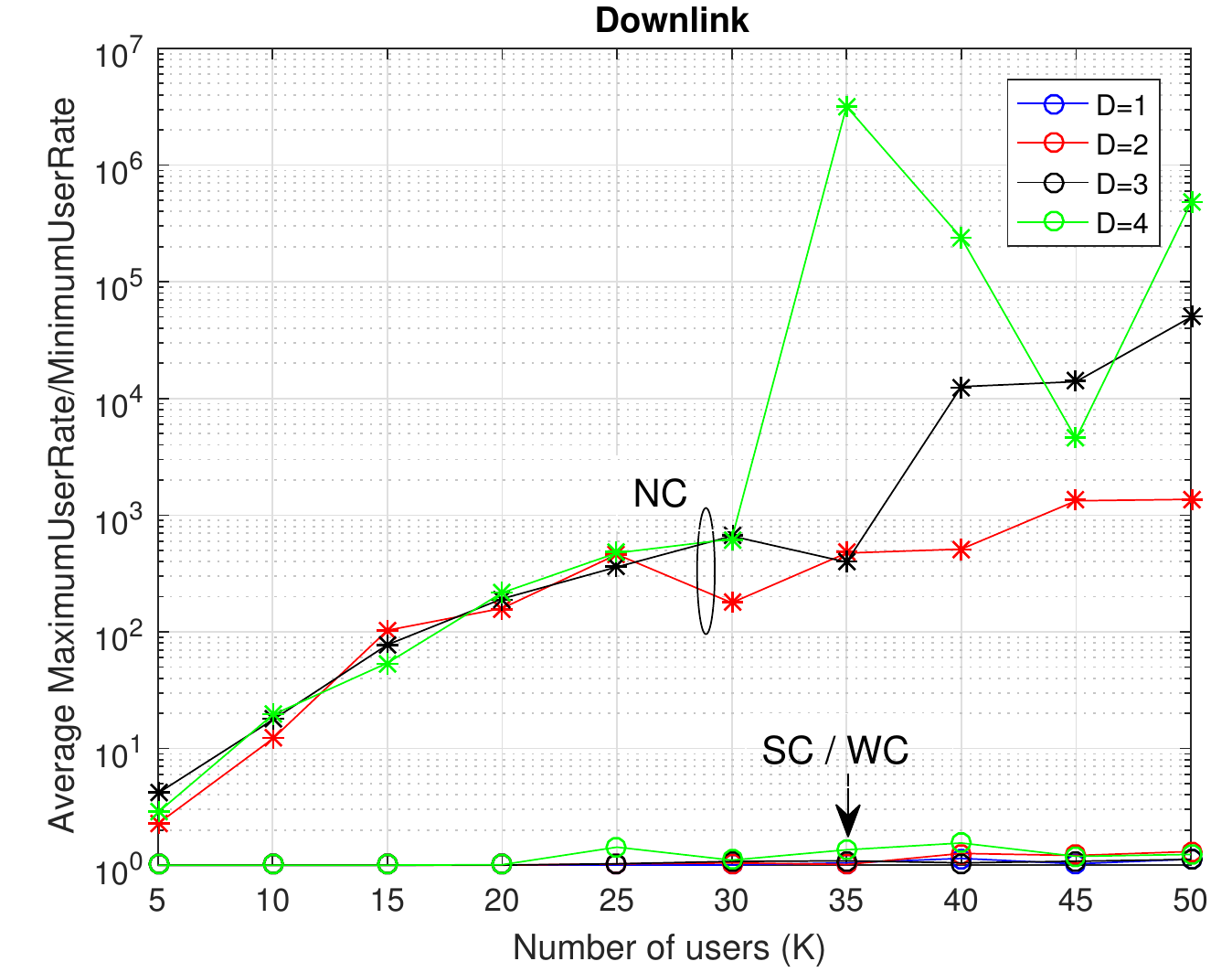}
    \caption{Average Maxrate/Minrate for SC, WC and NC.}
    \label{quotient}
\end{figure}

Figure~\ref{max-min} depicts the average max-min performance (i.e., average minimum user rate throughout the deployment area) as a function of the network load ($K$) for the different connectivity strategies and when using a varying number of CPUs. Recall that the SC strategy, regardless of the number of CPUs, is equivalent to a centralized scheme and can therefore be used as a performance upper bound (the figure only shows SC results for $D=1$). Indeed it can be observed how the SC outperforms any other scheme irrespective of the network load. Moving now to the WC results, it can be clearly seen that increasing the number of CPUs in the system leads to some degradation caused by the inter-cluster interference that cannot now be suppressed. Nevertheless, note that even with $D=4$, a rather moderate loss of roughly a 15-20\% is observed with respect to a fully centralized (or SC scheme) while the backhaul requirements to support this form of cooperation are rather modest in comparison to the SC approach. Turning now our attention to the NC strategy, it can be clearly appreciated how the independent optimization (in terms of power and pilot sequence allocation) has a very deleterious impact on the max-min performance. Specially dramatic is to observe how for a moderate-to-large number of users the minimum rate quickly approaches zero, a condition indicative that some users in the network have ceased to be served. In particular, the NC strategy cannot provide any acceptable minimum rate for $K>35$.

To further highlight the benefits of the WC strategy, Fig.~\ref{quotient} represents the average quotient between the maximum and minimum user rate for different number of users. It is well-known that, under the maxmin criterion, all users throughout the network should be found to have the same rate (except for numerical approximations when solving \eqref{optim}). This figure clearly shows that the WC connectivity level, despite sharing only large-scale information, is able, up to a large extent, to maintain the equal user rate property. In contrast, under the NC strategy and caused by its poorer interference control, user rates can differ vastly, often by various orders of magnitude with this tendency becoming far more apparent the more users that are active in the system.

\section{Conclusion}
This paper has analysed a key scalability issue when deploying CF-M-MIMO, namely, how to link various CPUs when the single-CPU CF-M-MIMO becomes difficult to implement in practice. Most importantly, this work has shown that simply allowing the CPUs to exchange large-scale CSI is enough to reap most of the benefits and performance of the single-CPU CF-M-MIMO, thus making the proposed decentralized architecture a suitable way to improve the practical scalability of the CF-M-MIMO concept. A novel distributed maxmin power allocation technique has been introduced that has been shown to preserve the provision of a uniform QoS throughout the network. Future work will progress by considering how different forms of clustering may help in further reducing the amount of CSI CPUs have to exchange and also by re-considering the problem in light of finite-capacity fronthauls and backhauls.

\end{document}